\documentclass[prb,showpacs,twocolumn,aps,superscriptaddress,a4paper]{revtex4-1}
\usepackage{dcolumn,amssymb,amsmath,amsfonts,graphicx,latexsym,color}

\begin{document}

\title{Effect of incommensurate disorder on the resonant tunneling through Majorana bound states on the topological superconductor chains}

\author{Pei Wang}
\email{wangpei@zjut.edu.cn}
\affiliation{Institute of Applied Physics, Zhejiang University of Technology, Hangzhou 310023, China}
\author{Shu Chen}
\email{schen@aphy.iphy.ac.cn}
\affiliation{Beijing National Laboratory for Condensed Matter Physics, Institute of Physics, Chinese Academy of Sciences, Beijing 100190, China}
\author{Gao Xianlong}
\email{gaoxl@zjnu.edu.cn}
\affiliation{Department of Physics, Zhejiang Normal University, Jinhua 321004, China}

\date{\today}

\begin{abstract}
We study the transport through the Kitaev chain with incommensurate potentials coupled to two normal leads by the numerical operator method. We find a quantized linear conductance of $e^2/h$, which is independent to the disorder strength and the gate voltage in a wide range, signaling the Majorana bound states. While the incommensurate disorder suppresses the current at finite voltage bias, and then narrows the linear response regime of the $I-V$ curve which exhibits two plateaus corresponding to the superconducting gap and the band edge respectively. The linear conductance abruptly drops to zero as the disorder strength reaches the critical value $2g_s+2\Delta$ with $\Delta$ the p-wave pairing amplitude and $g_s$ the hopping between neighbor sites, corresponding to the transition from the topological superconducting phase to the Anderson localized phase. Changing the gate voltage also causes an abrupt drop of the linear conductance by driving the chain into the topologically trivial superconducting 
phase, whose $I-V$ curve exhibits an exponential 
shape.
\end{abstract}

\maketitle

\section{Introduction}

It is well known that the resonant tunneling through a localized level results in a quantized conductance~\cite{jauho}, which was observed in the transport through quantum dots~\cite{wiel00}. Recently, researchers~\cite{law} found that the resonant tunneling with a quantized conductance also happens in the presence of Majorana bound states, providing a way of detecting the Majorana fermions~\cite{beenakker11}. From then on, there is a growing interest in the study of the quantum transport through the systems that accommodate Majorana bound states~\cite{bolech07,flensberg10,pikulin12,liu12,fregoso,tanaka04}.

The Majorana fermion is a proposed charge-neutral particle of its own antiparticle~\cite{beenakker11}, which continuously attracts the efforts of searching for its existence in nature~\cite{moore,rice,fu07,sau,nilsson,oreg10}, due to its potential application in topological quantum computation~\cite{kitaev00,nayak}. The Majorana fermions were predicted to emerge as the edge quasi-particle excitations of the topological superconductors~\cite{fu07,sau,nilsson,oreg10}. Some proposals have been introduced for observing the Majorana fermions in these systems, e.g., by the topological Josephson effect (see Ref.~\cite{Ioselevich,Nogueira,Beenakker} for recent progress). But no experiments realized these proposals up to now. Law {\it et al.}~\cite{law} proposed a method for detecting the Majorana bound states in topological superconductors by the effect of the resonant Andreev reflection across a junction between a normal metal and a topological superconductor, which contributes a quantized conductance of $2e^2/h$. 
Recent experimental results~\cite{mourik12,deng12,das12} are qualitatively consistent with the existence of Majorana bound states in these systems. However, the observed zero bias peak in experiments does not exclusively signal the Majorana bound states~\cite{pikulin12,
liu12,fregoso,lee13}, mainly because the experiments were conducted in real materials where the disorder is unavoidable and may induce additional zero bias peak. To better understand the experimental results, it is necessary to study the effect of disorder~\cite{asano} on the transport through topological superconductors.

Compared to the random disorder, the disorder produced by incommensurate potentials can induce a finite transition from extended states to Anderson localized states in one-dimensional incommensurate lattices~\cite{AA}. Due to its good controllability, the incommensurate potential has been experimentally engineered with ultracold atoms loaded in 1D bichromatic optical lattices~\cite{Roati}, simulating intensive study of the localization properties of quasi-periodic systems~\cite{tezuka10,tezuka}. Particularly, the effect of incommensurate potentials on the one-dimensional topological superconductors has also been theoretically studied~\cite{degottardi13a,cai13}. As a prototype model for describing the one-dimensional topological superconductors, the Kitaev's p-wave superconductor model has been widely used as an effective model in current theoretical studies from various aspects \cite{kitaev00,degottardi13a,cai13,Jiang}. Its realization in cold atom quantum wires is discussed by Jiang {\it et al.}~\cite{Jiang}
. Furthermore, Fulga {\it et al.}~\cite{fulga12} discussed the possibility of realizing the Kitaev chain in experiments by a chain of quantum dots coupled to superconductors. The incommensurate potentials can be produced by well tuning the gate voltages applied to different dots. Another different proposal is by using the quantum wires~\cite{oreg10}, while the incommensurate potentials can be generated electrically~\cite{gang12}.

The Kitaev's model~\cite{kitaev00} describes a p-wave topological superconductor. The recent study~\cite{degottardi13a,cai13} discovered that in the presence of incommensurate potentials the system undergoes a quantum phase transition from the superconducting phase to the Anderson localized phase as the strength of disorder increasing. This conclusion is supported by the energy spectra and the calculation of the topological invariant. However, the study of the transport through a junction containing the Kitaev chain with incommensurate potentials is still lack. In this paper, we employ the numerical operator method~\cite{wang13} to study this problem. Our results clarify the effect of incommensurate disorder on the transport through topological superconductors, and provide information of distinguishing the different phases by the feature of the current-voltage curves. Furthermore, the study of transport theory and experiment for transition from the metal to the localized state in one-dimensional setups has never been carried out before since the traditional one-dimensional Anderson model has no a finite transition. The transport analysis in our paper fills this gap.

The rest of paper is organized as follows. In Sec.~II, we briefly review the Kitaev's p-wave superconductor model with incommensurate potentials, introduce our junction model for studying the transport, and discuss the details of the numerical operator method. We present the results of the current-voltage curves and the linear conductance as the gate voltage is zero in Sec.~III, and those as the gate voltage deviates from zero in Sec.~IV. Sec.~V is a short summary.

\section{Model and method}

\subsection{Kitaev chain with incommensurate potentials}

The Hamiltonian describing the Kitaev chain with incommensurate potentials is written as
\begin{equation}\label{hamiltoniansc}
\begin{split}
  \hat H_{sc} = & -\sum_{i=0}^{L-2} g_s (\hat c^\dag_i \hat c_{i+1} + h.c.) + \Delta \sum_{i=0}^{L-2} (\hat c^\dag_i \hat c^\dag_{i+1} + h.c.). \\ & + \sum_{i=0}^{L-1} V_i \hat c^\dag_i \hat c_i,
\end{split}
\end{equation}
where $L$ denotes the length of the chain, $g_s$ the hopping amplitude, $\Delta$ the p-wave pairing amplitude induced by a RF field in cold atom systems~\cite{Jiang} or by the proximity effect in solid state systems~\cite{fulga12}, and $V_i=V_g + V_{d} \cos(2\pi i\alpha)$ the incommensurate potential varying at each lattice site. In the expression of $V_i$, $V_g$ denotes the gate voltage or the overall shift of the energy levels, $V_d$ the strength of disorder, and $\alpha$ an irrational number taken to be $\alpha=(\sqrt{5}-1)/2$ in the present study. This Hamiltonian~\cite{kitaev00} describes a p-wave topological superconductor supporting zero-energy Majorana bound states localized at the edges of the chain for $V_d < (2g_s + 2\Delta)$. As $V_d>(2g_s + 2\Delta)$ the system experiences a quantum phase transition~\cite{cai13} from the topological superconducting phase to the Anderson localized phase.

The Hamiltonian (\ref{hamiltoniansc}) can be diagonalized as
\begin{equation}
\hat{H}_{sc}= \sum_{n=1}^{L} \lambda_n \left( \hat \eta^\dag_n \hat \eta_n  -1/2 \right),
\end{equation}
 by introducing a Bogoliubov-de Gennes (BdG) transformation $\hat \eta_n = \sum_{i=0}^{L-1} \left( u_{n,i} \hat c_i + v_{n,i} \hat c_i^\dag\right)$, where $\lambda_n $ denotes the quasi-particle energy, and $u_{n,i}$ and $v_{n,i}$ can be obtained by numerically solving the corresponding BdG equations. In order to obtain the tunneling current in the following, we need to calculate the ground-state correlation functions $ \mathcal{C}^{0,0}_{i,j} = \langle \hat c_i \hat c_j \rangle$, $ \mathcal{C}^{0,1}_{i,j} = \langle \hat c_i \hat c^\dag_j \rangle$, $ \mathcal{C}^{1,0}_{i,j} = \langle \hat c^\dag_i \hat c_j \rangle$ and $ \mathcal{C}^{1,1}_{i,j} = \langle \hat c^\dag_i \hat c^\dag_j \rangle$ (see the appendix~\ref{appendix:corr} for more details), written in terms of the amplitude of the BdG transformation as
\begin{equation}
 \begin{array}{cc}\displaystyle
\mathcal{C}^{0,0}_{i,j} = \sum_{n=1}^L v_{n,i} u_{n,j}, & \displaystyle \mathcal{C}^{0,1}_{i,j} = \sum_{n=1}^L v_{n,i} v_{n,j} , \\ \displaystyle \mathcal{C}^{1,0}_{i,j}=\sum_{n=1}^L u_{n,i} u_{n,j}, & \displaystyle  \mathcal{C}^{1,1}_{i,j} = \sum_{n=1}^L u_{n,i} v_{n,j}.
\end{array}
\end{equation}

\subsection{Transport through the Kitaev chain}

\begin{figure}
\includegraphics[width=1.0\linewidth]{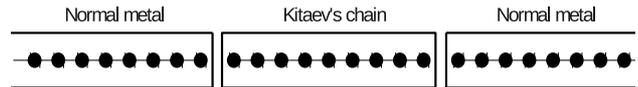}
\caption{The schematic diagram of the transport setup. Two normal leads are connected to the edge sites of a Kitaev chain.}\label{fig:schematic}
\end{figure}
To study the transport through the Kitaev chain with incommensurate potentials, we locate it between two single-channel normal leads (the N-S-N structure has been used with random disorder~\cite{fregoso}), described by the Hamiltonians
\begin{equation}
 \hat H_L = -g_l \sum_{i=-\infty}^{-2} (\hat c^\dag_i \hat c_{i+1} + h.c.)
\end{equation}
and
\begin{equation}
\hat H_R = - g_l \sum_{i=L}^\infty  (\hat c^\dag_i \hat c_{i+1} + h.c.)
\end{equation}
respectively, where $g_l$ is the hopping amplitude in the leads. The leads $L$ and $R$ are coupled to the left and right edge sites of the Kitaev chain respectively (see Fig.~\ref{fig:schematic} for the schematic diagram of the setup). The Hamiltonian describing the coupling between the leads and the chain is expressed as
\begin{equation}
 \hat H_V = g_c (\hat c^\dag_{-1} \hat c_{0} + \hat c^\dag_{L-1} \hat c_L + h.c.),
\end{equation}
where $g_c$ is the coupling strength. Then the total Hamiltonian of the transport setup is
\begin{equation}
 \hat H = \hat H_{sc} + \hat H_L + \hat H_R + \hat H_V,
\end{equation}
where $\hat H_{sc}$ is described by Eq.~(\ref{hamiltoniansc}).

We study the tunneling current from the left lead to the chain, expressed as
\begin{equation}\label{leftcdef}
\begin{split}
 I_L (t) =& -\langle \frac{d \hat N_L}{dt} \rangle \\
= &  -2 g_c \textbf{Im} \langle \hat c^\dag_{-1}(t) \hat c_0(t) \rangle ,
\end{split}
\end{equation}
and that from the chain to the right lead, expressed as
\begin{equation}\label{rightcdef}
\begin{split}
 I_R (t) =& \langle \frac{d \hat N_R}{dt} \rangle \\
= &  -2 g_c \textbf{Im} \langle \hat c^\dag_{L-1}(t) \hat c_L(t) \rangle.
\end{split}
\end{equation}

In this paper, the two leads are in different chemical potentials, being $\mu_L=V/2$ and $\mu_R=-V/2$ respectively, where $V$ is the voltage bias. We take the wide band limit in the leads and the intermediate coupling strength between the leads and the chain by setting $g_l=10g_s$ and $g_c=g_s$.

\subsection{The numerical operator method}

The numerical operator method~\cite{wang13} is employed to calculate $I_L(t)$ and $I_R(t)$. For simplicity, we define the synonyms $\hat d_{j1}=\hat c^\dag_j$ and $\hat d_{j0} = \hat c_j$ in this subsection.

We first solve the Heisenberg equation $\frac{d}{dt} \hat d_{js}(t) = i[\hat H, \hat d_{js} (t)] $ to obtain the expressions of $\hat c^\dag_{-1}(t)$, $\hat c_0(t)$, $\hat c^\dag_{L-1}(t)$ and $\hat c_L(t)$. Since $\hat H$ is quadratic, the solution is supposed to be
\begin{equation}\label{heisenbergsolution}
 \hat d_{js} (t) = \sum_{ks'} W_{js,ks'}(t) \hat d_{k s'}.
\end{equation}
Analytically solving the propagators $W_{js,ks'}(t)$ is difficult in the presence of disorder, however, they can be obtained through the numerical calculation iteratively. From $\hat d_{js} (t+\Delta t)=e^{i\hat H \Delta t} \hat d_{js}(t) e^{-i\hat H \Delta t} $ and Eq.~(\ref{heisenbergsolution}), we have
\begin{equation}\label{iterativerelation}
\begin{split}
\sum_{ks'} W_{js,ks'}(t+\Delta t) & \hat d_{ks'} \\
= \sum_{ks'} W_{js,ks'}(t) & \bigg( \hat d_{ks'}+ i\Delta t [\hat H, \hat d_{ks'}] \\ &+ \frac{(i\Delta t)^2}{2} [\hat H,[\hat H,\hat d_{ks'}]] + O(\Delta t^3) \bigg),
\end{split}
\end{equation}
where the time step $\Delta t$ is set to be small enough so that the terms $O(\Delta t^3)$ can be neglected. $W_{js,ks'}(t+\Delta t)$ can then be expressed as a linear function of $W_{js,ks'}(t)$:
\begin{equation}
\begin{split}
 W_{js,ks'}(t+\Delta t)= & W_{js,ks'}(t) +i\Delta t \sum_{ls''} W_{js,ls''}(t) G_{ls'',ks'} \\ & -\frac{\Delta t^2}{2} \sum_{ls_1,ms_2} W_{js,ls_1}(t) G_{ls_1,ms_2} G_{ms_2,ks'}  ,
\end{split}
\end{equation}
where the coefficients $G_{ks_1,ls_2}$ are defined by the commutator $[\hat H, \hat d_{ks_1}]= \sum_{ls_2} G_{ks_1,ls_2}\hat d_{ls_2} $. We work out the propagator $W_{js,ks'}(t)$ at an arbitrary time by an iterative algorithm starting from $t=0$ when $W_{js,ks'}(0)= \delta_{j,k}\delta_{s,s'}$ and moving forward $\Delta t$ at each step. The error caused by a finite $\Delta t$ is in the order $O(\Delta t^3)$ in our calculation. One can keep the higher order terms $O(\Delta t^n)$ with $n>2$ in Eq.~(\ref{iterativerelation}) so as to reduce the error to the order $O(\Delta t^{n+1})$. This makes it possible to use a larger $\Delta t$ without reducing the precision of the result. But, in practice, we find that keeping to the order $O(\Delta t^2)$ is efficient enough for our purpose.

At each step, we store all the non-zero propagators $W_{js,ks'}(t)$ and use them to calculate the propagators at next step. The number of non-zero propagators increases with the step quickly. So we apply a truncation scheme to keep only a fixed number of non-zero propagators (the number is denoted by $M$) with the largest magnitudes at each step. This truncation scheme is necessary for keeping the computing resource within a reasonable amount. The value of $M$ is determined empirically. In this paper, setting $M$ to several tens of thousands is enough for obtaining the current with the relative error below $10^{-5}$.

Then we substitute the solution in Eq.~(\ref{heisenbergsolution}) into Eqs.~(\ref{leftcdef}) and~(\ref{rightcdef}), and express the currents as
\begin{equation}\label{expcurrent}
\begin{split}
 I_L(t) & = -2 g_c \sum_{j,s,j',s'}\textbf{Im} \left[ W_{-11,js}(t) W_{00,j's'}(t)\right] \langle \hat d_{js} \hat d_{j's'} \rangle, \\
 I_R(t) & = -2 g_c \sum_{j,s,j',s'}\textbf{Im} \left[ W_{L-1,1,js}(t) W_{L0,j's'}(t)\right] \langle \hat d_{js} \hat d_{j's'} \rangle.
\end{split}
\end{equation}
Here the correlation function is $\langle \hat d_{js} \hat d_{j's'} \rangle=\mathcal{C}^{s,s'}_{j,j'}$ as $0\leq j,j' \leq L-1$. And as $j$ and $j'$ are both in the lead $L$ or $R$, we have
\begin{eqnarray}
\left\{ \begin{array}{c} \langle \hat d_{j1} \hat d_{j'0} \rangle = \displaystyle \frac{\sin((j-j')\theta )}{\pi(j-j')} \\ \langle \hat d_{j0} \hat d_{j'1} \rangle = \delta_{j,j'}-\displaystyle \frac{\sin((j-j')\theta )}{\pi(j-j')} \end{array} \right.
\end{eqnarray}
with $\theta=\arccos(-V/4g_l)$ as $j,j'<0$ or $\theta=\arccos(V/4g_l)$ as $j,j'\geq L$. The correlation function is zero in all the other cases.

To demonstrate the power of our method, we compare it with exact diagonalization at $V_g=V_d=0$ (see more details in appendix~\ref{appendix:comp}). The comparison shows that, for the long relaxation time of currents, our method gives better performance than the exact diagonalization method.

\begin{figure}
\includegraphics[width=1.0\linewidth]{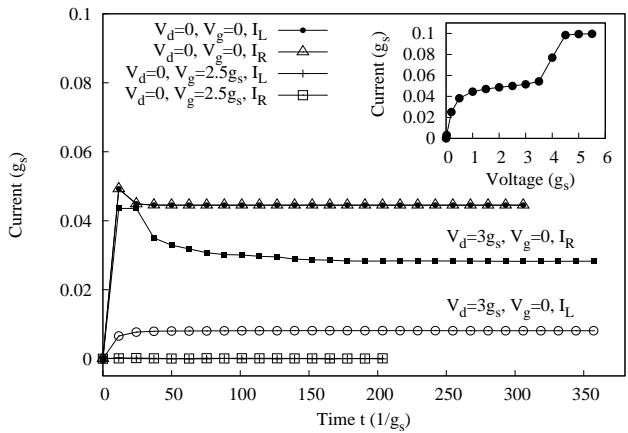}
\caption{The evolution of the currents after the coupling between the leads and the Kitaev chain is switched on at the time $t=0$. The p-wave pairing amplitude and the voltage bias are set to $\Delta=V=g_s$. We compare the left and right currents $I_L(t)$ and $I_R(t)$ at different disorder strength $V_d$ and gate voltage $V_g$, represented by different types of points. The solid lines are the guide for the eyes. The left and right currents coincide well with each other except for $V_d=3g_s$ and $V_g=0$. [Inset] The $I-V$ curve obtained at $\Delta=g_s$ and $V_d=V_g=0$, in which case there is a single $I-V$ curve since $I_L=I_R$. }\label{fig:evolution}
\end{figure}
What are interesting in experiments are the stationary currents $I_L=\lim_{t\to \infty} I_L(t)$ and $I_R=\lim_{t\to\infty}I_R(t)$. In practice, we obtain $I_L$ and $I_R$ from $I_L(t)$ and $I_R(t)$ respectively at the times much larger than the current relaxation time. As shown in Fig.~\ref{fig:evolution}, $I_L(t)$ and $I_R(t)$ quickly relax to their stationary values within a few oscillations in the absence of disorder. While at a large disorder strength, the current relaxes slowly, since the disorder blocks the movement of electrons (see the curves titled $V_d=3g_s$ in Fig.~\ref{fig:evolution}). 

We obtain the stationary currents $I_L$ and $I_R$ at different voltage biases and plot the corresponding current-voltage ($I-V$) curves (see the inset of Fig.~\ref{fig:evolution} for an example). The linear conductance is the slope of the $I-V$ curve at $V=0$, and in practice is got by calculating the ratio of the current to the voltage bias at small $V$ in the linear response regime:
\begin{equation}
 G = \lim_{V\to 0} \frac{I_\alpha}{V},
\end{equation}
where $\alpha=L,R$.

\section{Transport at zero gate voltage}

\subsection{The resonant tunneling via Majorana bound states}

\begin{figure}
\includegraphics[width=1.0\linewidth]{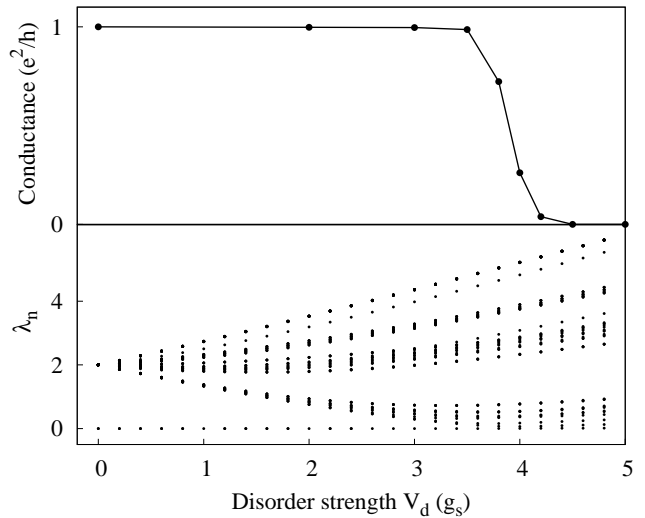}
\caption{[Top panel] The conductance $I_L/V=I_R/V$ at $V=0.005g_s$ as a function of the disorder strength $V_d$, read out from the $I-V$ curves (for the cases of given $V_d$ see Fig.~\ref{fig:ivcurve}). The p-wave pairing amplitude is $\Delta=g_s$ and the gate voltage is $V_g=0$. This conductance is just the linear conductance except for the points close to the critical point $V_d=4g_s$, where the linear response regime is smaller than $V=0.005g_s$. [Bottom panel] Energy spectra of the Kitaev chain at the same parameters under the open boundary condition. Only the quasi-particle energies $\lambda_n\geq 0$ are shown due to the particle-hole symmetry. The energy gap closes at the point $V_d=4g_s$ (not clearly seen in the figure because the gap close to the transition point is too small), coinciding with the result in Ref.~\cite{cai13}}\label{fig:transition}
\end{figure}
As $V_d <2g_s+2\Delta$, the Kitaev chain is in the superconducting phase. The voltage drops only at the interfaces between the chain and the leads, but not inside the chain. The chemical potential of the chain is fixed to $(\mu_L + \mu_R)/2=0$, i.e., the chain is grounded. The pair of Majorana bound states locates exactly at zero energy, inside the transmission window $[-V/2,V/2]$. So we observe the linear conductance of $e^2/h$ independent to the disorder strength $V_d$ (see the top panel of Fig.~\ref{fig:transition}), the same conductance as that of the resonant tunneling through a single level without the spin degrees of freedom. The linear conductance disregarding the disorder is protected by the superconducting gap, which isolated the Majorana bound states from those in the bands (see the energy spectra in the bottom panel of Fig.~\ref{fig:transition}). In the transmission window, there are only a pair of Majorana states left, which locate at the two edges of the chain, coupled strongly to the left 
and 
right leads. The Majorana states participate in shuttling the electrons, resulting in the quantized conductance, which signals the resonant tunneling via the Majorana bound states.

Law et al. predicted in 2009 that the linear conductance of the junction between a normal lead and a topological superconductor is $2e^2/h$ due to the resonant Andreev reflection~\cite{law}. Our result coincides exactly with theirs. The current goes across two junctions in our model, each of which contributes a resistance of $h/(2e^2)$ according to Law et al., so that the total conductance of the two parallel-connected junctions is exactly $e^2/h$.

Furthermore, a quantized conductance excludes the possibility of the chain's being in the normal metal phase. In Fig.~\ref{fig:ivcurve}, we show the current-voltage curves of the system in different phases. The linear conductance is the slope of the $I-V$ curve at the origin. We see that the linear conductance of the normal metal phase ($\Delta=V_d=0$) is obviously lower than that of the topological superconducting phase ($\Delta=g_s, V_d=0$). Even a chain in the normal metal phase also has scattering states around zero energy, but these states are not localized at the edges. In fact, the coupling strength of these states to the leads scales as $1/\sqrt{L}$ where $L$ denotes the chain length, becoming infinitesimal in the thermodynamic limit, while, the Majorana states locate always at the edges and couple strongly to the leads, independent to the size of the chain. So the resonant tunneling can happen via the Majorana states, but not via the scattering states of the normal metal phase.

\subsection{The transition from the topological superconducting phase to the Anderson localized phase}

\begin{figure}
\includegraphics[width=1.0\linewidth]{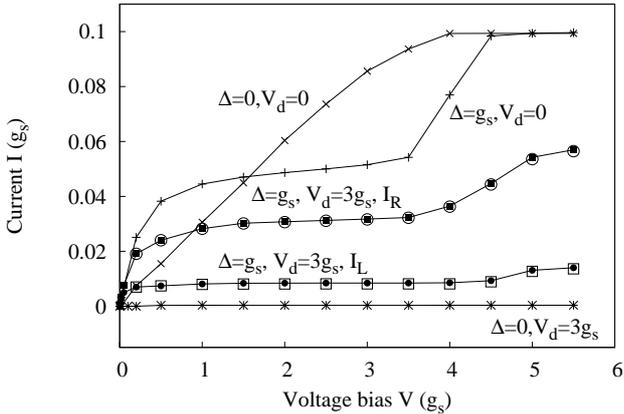}
\caption{We plot the $I-V$ curves in the clean topological superconducting phase at $\Delta=g_s,V_d=0$ (daggers) , in the disordered topological superconducting phase at $\Delta=g_s,V_d=3g_s$ with $I_L$ (full dots) and $I_R$ (full squares) separately representing the left and right currents, in the normal metal phase at $\Delta=0,V_d=0$ (crosses), and in the Anderson localized phase at $\Delta=0,V_d=3g_s$ (stars). At $\Delta=g_s,V_d=3g_s$, we also rotate the chain by $180^\circ$ and reconnect it to the leads, and plot the corresponding left (empty circles) and right currents (empty squares) respectively. The solid lines are the guide for the eyes. The left and right currents are the same, except for $\Delta=g_s,V_d=3g_s$. At $\Delta=g_s,V_d=3g_s$, the left and right currents exchange with each other as we rotate the chain by $180^\circ$, which is shown by the fact that the full squares (dots) fit well with the empty circles (squares).}\label{fig:ivcurve}
\end{figure}
As the disorder strength $V_d$ increases but keeps in the range $(0,2g_s+2\Delta)$, the linear conductance keeps invariant. But as $V_d > 2g_s+2\Delta$, the chain is driven into the Anderson localized phase, in which all the quasi-particle states are localized and then the chain is insulating. As a result, the linear conductance abruptly drops to zero as $V_d$ crosses the critical point $2g_s+2\Delta$, as shown in the top panel of Fig.~\ref{fig:transition}.

We observe the transition from the superconducting phase to the insulating phase not only in the linear conductance but also in the energy spectra. In the bottom panel of Fig.~\ref{fig:transition}, we see that the superconducting gap gradually decreases as $V_d$ increasing. The gap totally closes at the critical point, indicating the phase transition, and then reopens again. The linear conductance keeps $e^2/h$ before the gap closes. But beyond the critical point, the zero-energy modes (the Majorana bound states) diminish, and the localized states open an energy gap~\cite{smallgap}. The diminish of the zero-energy modes leads to a zero linear conductance.

In Fig.~\ref{fig:transition}, we see that the drop of the linear conductance happens in the range $[3.5g_s,4.5g_s]$, but not exactly at the critical point $V_d=4g_s$ as expected by the phase transition theory. We ascribe this smooth transition to two reasons. First, the Kitaev chain in our calculation is finite with a length of $L=50$, but the nonanalytic behavior happens only in the thermodynamic limit. Second, in our calculation we take the conductance at small voltage bias ($V=0.005g_s$ in Fig.~\ref{fig:transition}) to be approximately the linear conductance. Around the critical point, the linear response regime is very small (we do observe the shrink of the linear response regime in Fig.~\ref{fig:ivcurve} as the disorder strength is closer to the critical point), so that the voltage bias that we choose is beyond the linear response regime.

\subsection{The current-voltage curves of different phases}

We compare the current-voltage curves of the normal metal phase, the topological superconducting phase and the Anderson localized phase in Fig.~\ref{fig:ivcurve}. Beyond the linear response regime, the $I-V$ curves also show abundant features that distinguish different phases.

In the normal metal phase, the current increases linearly with the voltage bias in a wide range, and then turns into a plateau at the point $V=4g_s$, which is just the width of the conduction band of the chain. The plateau indicates the limit of the current that the chain can carry. The linear response regime of the topological superconducting phase (the curve titled $\Delta=g_s,V_d=0$ in Fig.~\ref{fig:ivcurve}) is thinner than that of the normal metal phase. In the topological superconducting phase, as the voltage bias increasing, the current quickly runs into the first plateau associated with the superconducting gap. Within the superconducting gap there lie only a pair of Majorana bound states exactly at zero energy, then the current will not increase any more as the voltage bias is beyond the width of the zero-mode levels since no new levels enter into the transmission window. Around $V=4g_s$, the current increases again and runs into the second plateau coinciding with that in the normal metal phase, due 
to the participation of the bulk states in the electron transport (see the energy spectrum at $\Delta=g_s$ and $V_d=0$ in Fig.~\ref{fig:transition}).

As the disorder strength increasing to $V_d=3g_s$, the linear conductance (the slope of the $I-V$ curve at the origin) keeps invariant, but the current at finite voltage is suppressed. The two-plateau structure is also clear in the $I-V$ curve, but their heights are significantly smaller. Especially, the reduction of the current at the first plateau reflects the shrink of the linear response regime and then the width of the zero-mode levels. The disorder narrows the linear response regime, indicating that purifying the sample can improve the temperature for observing the quantized conductance, which is comparable with the width of the zero-mode levels.

In the presence of disorder, the left and right currents deviate from each other, even their slopes at the origin keep the same. The reason of $I_L\neq I_R$ at finite voltage is that the p-wave pairing interaction breaks the conservation of the electron number ($I_L=I_R$ is guaranteed at $\Delta=0$). At the same time, the incommensurate potentials break the left-right symmetry of the Kitaev chain. If rotating the chain by $180^\circ$ and then connecting it to the leads again, we see that the left and right currents exchange with each other (see Fig.~\ref{fig:ivcurve}).

In the Anderson localized phase, the chain is insulating with all the quasi-particle eigenstates localized. At $\Delta=0$ and $V_d=3g_s$, we see that the current is approximately zero in the whole range of voltage bias, reflecting the lack of a scattering state which can shuttle the electrons from the left lead to the right lead.

\section{Transport at a biased gate voltage}

\begin{figure}
\includegraphics[width=1.0\linewidth]{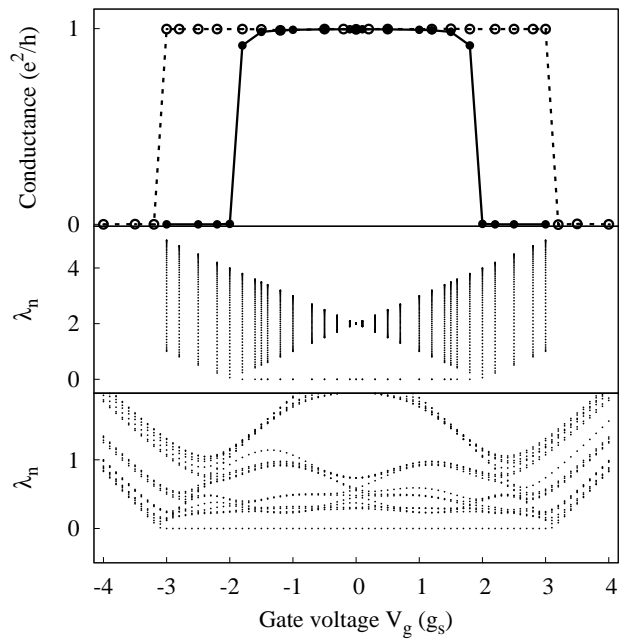}
\caption{[Top panel] The conductance $I_L/V=I_R/V$ as a function of the gate voltage at $\Delta=g_s,V_d=0$ and $V=0.02g_s$ (empty circles) and at $\Delta=g_s,V_d=3g_s$ and $V=0.005g_s$ (solid circles). The lines are the guides for the eyes. In both cases, the sharp drop of the conductance happens at the critical point dividing the topological superconducting phase and the topologically trivial superconducting phase. [Middle panel] Energy spectra of the Kitaev chain at $\Delta=g_s,V_d=0$ under the open boundary condition. [Bottom panel] Energy spectra of the Kitaev chain at $\Delta=g_s,V_d=3g_s$ under the open boundary condition. }\label{fig:gatevoltage}
\end{figure}
An important feature of the Majorana state in the topological superconducting systems is that its energy is fixed to zero. In contrast, the energy of the quasi-particle states in the topologically trivial phase can be adjusted by the gate voltage applied to the system. For example, in the transport through a quantum dot in the Coulomb blockade regime, the linear conductance depends on the gate voltage and shows a series of Lorentzian peaks. However, in our model, the energy of the Majorana state is fixed inside the transmission window $[-V/2,V/2]$, then the resonant tunneling through it should not be affected by the gate voltage to the chain. This feature distinguishes the transport through the topological superconducting phase from that through the topologically trivial phase without zero modes.

To verify this feature of the Majorana state, we apply a non-zero gate voltage $V_g$, and find that the linear conductance does keep the quantized value $e^2/h$ disregarding the gate voltage as it is in a certain range. But the linear conductance abruptly drops to zero as $V_g$ is outside this range. In the top panel of Fig.~\ref{fig:gatevoltage}, we see that the conductance plateau is in the range $V_g\in [-2g_s,2g_s]$ at $V_d=0$, while in the range $V_g\in[-3g_s,3g_s]$ at $V_d=3g_s$, as $\Delta=g_s$.

The plateau structure of the conductance function in fact reflects the quantum phase transition from the topological superconducting phase to the topologically trivial superconducting phase. In the middle and bottom panels of Fig.~\ref{fig:gatevoltage}, we show the energy spectra of the Kitaev chain varying with $V_g$ at $V_d=0$ and $V_d=3g_s$ respectively. The common feature of the energy spectra is that there exists the zero modes accompanied by a finite energy gap in a range of $V_g$. In this range, the chain is a topological superconductor with a pair of Majorana bound states supporting the resonant tunneling. The energy gap closes at the boundary of the topological superconducting phase, and then reopens as $V_g$ is outside the topological superconducting phase, accompanied by the diminish of the zero modes which results in a zero linear conductance. In Fig.~\ref{fig:gatevoltage}, we see that the points where the energy gap closes are $V_g =\pm 2g_s$ for $V_d=0$ and $V_g=\pm 3g_s$ for $V_d=3g_s$, 
coinciding well with the critical points where the conductance changes.

\begin{figure}
\includegraphics[width=1.0\linewidth]{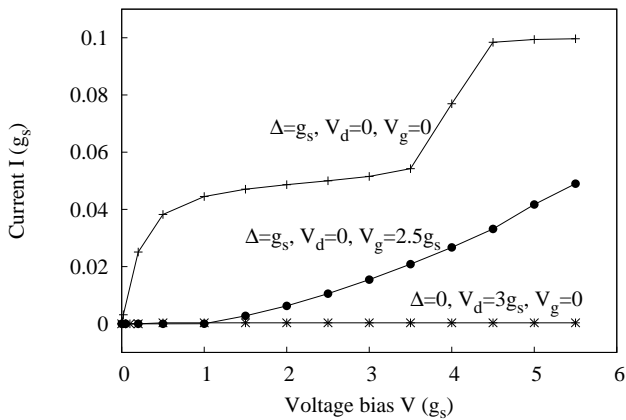}
\caption{The $I-V$ curves in the topological superconducting phase (labeled by $\Delta=g_s,V_d=0,V_g=0$), in the topologically trivial superconducting phase (labeled by $\Delta=g_s,V_d=0,V_g=2.5g_s$), and in the Anderson localized phase (labeled by $\Delta=0,V_d=3g_s,V_g=0$).}\label{fig:gateivcurve}
\end{figure}
And we signal the topologically trivial superconducting phase also by its $I-V$ curve, which exhibits an exponential shape (see Fig.~\ref{fig:gateivcurve}). The exponential $I-V$ curve always indicates a gap in the system, which is in fact the superconducting gap. The shape of the $I-V$ curve in the trivial superconducting phase is quite different from that in the Anderson localized phase, even the linear conductances in the two phases are both zero. In the superconducting phase, the chain can carry a strong current as the voltage bias overcomes the gap. But in the Anderson localized phase, the current keeps zero even at a large bias, because all the states are localized and cannot carry the current.

\section{Conclusions}

In this paper, we employ the numerical operator method to study the transport through the Kitaev chain with incommensurate potentials located between two normal leads. We find a quantized linear conductance of $e^2/h$ as the chain is in the topological superconducting phase, independent to the disorder strength or the gate voltage applied to the chain. By studying the energy spectra of the chain, we find that the quantized conductance results from the resonant tunneling via the Majorana bound states lying exactly at the center of the transmission window and protected by an energy gap. While increasing the disorder strength or changing the gate voltage, at some critical point, the linear conductance abruptly drops to zero, corresponding to the transition of the chain into the Anderson localized phase or the topologically trivial superconducting phase respectively. At the transition point, the energy gap closes, and then reopens with the zero modes diminishing. The different phases can be distinguished by 
their $I-V$ curves. The $I-V$ curve of the normal metal phase has a single plateau, while that of the topological superconducting phase has two, corresponding to the superconducting gap and the band edge respectively. The linear response regime of the topological superconducting phase is small compared to that of the normal metal phase, and is further shallowed by the disorder. The $I-V$ curves of the Anderson localized phase and the topologically trivial phase both have a zero slope at the origin. However, the latter exhibits an exponential shape signaling the superconducting gap, while the former is approximately zero in the whole range of voltage because all the quasi-particle states are localized.

\section*{Acknowledgements}
\label{acknowledgements}
P. Wang has been supported by the NSF of China under Grant No.11304280. S. Chen has been supported by the NSF of China under Grants No.11174360 and No.11121063. And Gao X. was supported by the NSF of China under Grants Nos. 11374266 and 11174253 and by the Zhejiang Provincial Natural Science Foundation under Grant No. R6110175.

\begin{appendix}
\section{The calculation of the correlation functions and the energy spectra in the Kitaev chain}
\label{appendix:corr}

To calculate the ground-state properties of the Kitaev chain, we re-express the Hamiltonian~(\ref{hamiltoniansc}) in a matrix form:
\begin{equation}
 \hat H_{sc} = \frac{1}{2} \left( \hat c^\dag_0 \cdots \hat c^\dag_{L-1} \hat c_0 \cdots \hat c_{L-1} \right)  \mathcal{H}_{BDG} \left(\begin{array}{c}  \hat c_0 \\ \vdots \\ \hat c_{L-1} \\ \hat c^\dag_0 \\ \vdots \\ \hat c^\dag_{L-1}  \end{array} \right),
\end{equation}
where the Bogoliubov-de Gennes matrix is written in blocks as $\mathcal{H}_{BDG} =  \left( \begin{array}{cc} \hat{h} & \hat{\Delta} \\ -\hat{\Delta} & -\hat{h} \end{array} \right)$ with $\hat h_{i,j} = \delta_{i,j} V_i - g_s\delta_{i,j+1} -g_s\delta_{i,j-1} $ and $\hat{\Delta}_{i,j} = \delta_{i,j-1}\Delta - \delta_{i,j+1} \Delta$.

We diagonalize the Bogoliubov-de Gennes matrix, and re-express the Hamiltonian into $\hat{H}_{sc}= \sum_{n=1}^{L} \lambda_n \left( \hat \eta^\dag_n \hat \eta_n  -1/2 \right)$, where $\hat \eta_n = \sum_{i=0}^{L-1} \left( u_{n,i} \hat c_i + v_{n,i} \hat c_i^\dag\right)$ and $\lambda_n \leq 0 $ denotes the quasi-particle energy in the negative energy branch. The energy spectrum $\lambda_n$ and the coefficients $u_{n,i}$ and $v_{n,i}$ satisfy the eigen equation
\begin{equation}
\mathcal{H}_{BDG} \left( \begin{array}{c} u_{n,0} \\ \vdots \\u_{n,L-1} \\ v_{n,0} \\ \vdots \\ v_{n,L-1} \end{array} \right) = \lambda_n \left( \begin{array}{c} u_{n,0} \\ \vdots \\u_{n,L-1} \\ v_{n,0} \\ \vdots \\ v_{n,L-1} \end{array} \right),
\end{equation}
and are determined by a numerical diagonalization routine.

In the ground state, the negative single-particle levels are all occupied while the positive ones are empty, so that $\langle \hat \eta_n^\dag \hat\eta_{n'} \rangle = \delta_{n,n'}$ as $1\leq n,n' \leq L$. We then obtain the four types of correlation functions by using the inverse Bogoliubov-de Gennes transformation $\hat c_i = \sum_{n=1}^L \left( u_{n,i} \hat \eta_n + v_{n,i} \hat \eta_n^\dag \right)$. The correlation functions are expressed as
\begin{equation}
 \begin{array}{cc}\displaystyle
\mathcal{C}^{0,0}_{i,j} = \sum_{n=1}^L v_{n,i} u_{n,j}, & \displaystyle \mathcal{C}^{0,1}_{i,j} = \sum_{n=1}^L v_{n,i} v_{n,j} , \\ \displaystyle \mathcal{C}^{1,0}_{i,j}=\sum_{n=1}^L u_{n,i} u_{n,j}, & \displaystyle  \mathcal{C}^{1,1}_{i,j} = \sum_{n=1}^L u_{n,i} v_{n,j}.
\end{array}
\end{equation}

\section{The comparison between exact diagonalization and numerical operator method}
\label{appendix:comp}

\begin{figure}
\includegraphics[width=1.0\linewidth]{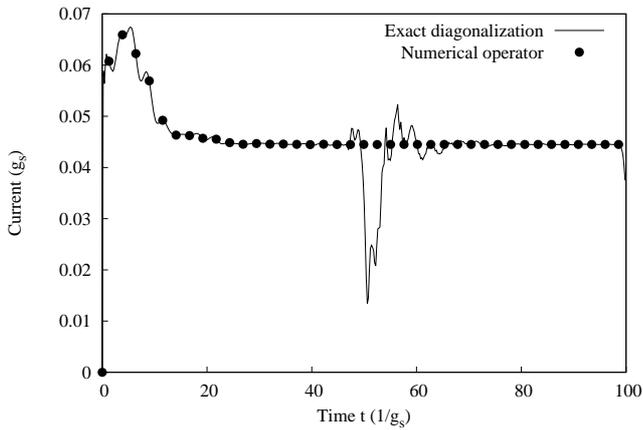}
\caption{The comparison between exact diagonalization and numerical operator method. We choose $V_d=V_g=0$, $\Delta=g_s$ and $V=g_s$. The left and right lead currents are the same in this case.}\label{fig:comparison}
\end{figure}
To demonstrate the power of our method, we compare the result by the numerical operator method with that by the exact diagonalization method. In exact diagonalization, we take the length of each normal lead to be $1000$. The currents by two different methods are plotted in Fig.~\ref{fig:comparison}. It is clear that the current by numerical operator method coincides very well with that by exact diagonalization in short time period. However, the current by exact diagonalization starts to oscillate as time increasing, the so-called finite size problem. While the {\it I-t} curve by numerical operator method still keeps smooth. The numerical operator method conquers the finite size problem by setting two infinite leads at very beginning, and is especially appropriate for studying the currents in the presence of disorder when the long relaxation time of currents (see Fig.~\ref{fig:evolution}) prevents exact diagonalization from reaching the stationary currents.

\end{appendix}


\begin{references}
\bibitem{smallgap} The gap in the bottom panel of Fig.~\ref{fig:transition} is quite small and does not show itself clearly, but the lowest levels in the Anderson localized phase are in fact finite. See the details in Ref.~[25].
\end{references}
\end{document}